\documentclass[a4paper]{article}
\usepackage{epsfig}

\def\be{\begin{equation}}
\def\ee{\end{equation}}
\def\bea{\begin{eqnarray}}
\def\eea{\end{eqnarray}}

\title{ Bose-Einstein condensation in the canonical ensemble}
\author{Daniel Bedingham}

\author{D.~J.~Bedingham\thanks{email:
{\tt d.j.bedingham@sussex.ac.uk}}  \\
{\small\it Centre for Theoretical Physics, University of Sussex,}\\
{\small\it Falmer, Brighton BN1 9QJ, United Kingdom.}
}

\date{\today}

\begin{document}

\maketitle

\begin{abstract}
Large-volume, high-temperature Bose-Einstein condensation is
illustrated for a relativistic $O(2)$-invariant scalar field with
fixed charge using the canonical ensemble. The standard, grand
canonical results are reproduced for the infinite-volume limit.
Finite-volume corrections are calculated in the canonical ensemble
and the results are found to differ from the finite-volume grand
canonical approximation in a consistent qualitative way.


\end{abstract}

\section{Introduction}

When dealing with conserved charges in a system in equilibrium the
usual procedure is to employ the grand canonical ensemble
\cite{h&w2,kap,ben}. The physical picture from which the grand
canonical ensemble is derived is that of a system in contact with
an infinite reservoir of particles with an associated chemical
potential for flow of particles into and out-of the system.
Averaging in the grand canonical ensemble is performed over all
possible charge states. The average charge is then fixed by a
definite choice of chemical potential.

By contrast, in the canonical ensemble, charge cannot move into
and out-of the system. There is no chemical potential. The charge
is fixed and any averaging must be performed only over states with
this definite fixed charge. This provides a more reasonable model
for a system which is insulated with respect to charge and a
better approximation for experiments involving Bose-Einstein
condensation (BEC) of trapped atoms \cite{BEC}.

A clear outline of the path integral formulation of relativistic
field theory with exactly conserved energy and charge can be found
in Refs.\cite{chai1,chai2}. The central technique is to insert
delta-functions into the trace over states in order to select only
those states with chosen eigenvalues of the Hamiltonian and charge
operators. This is the technique we shall use to exactly conserve
charge in the canonical ensemble. An example of the application of
this method to BEC in the canonical ensemble can be found in
Ref.\cite{poli} where fluctuations in the ground state occupation
are studied in the non-interacting, non-relativistic limit. A
further study of the comparison between canonical and grand
canonical ensembles with regard to trapped atoms can be found in
Ref.\cite{herz}.

In this article, we outline a fully relativistic calculation of
BEC using the canonical ensemble. After mathematically formulating
the model in the next section we go on to consider the
high-temperature, large-volume limit in section~\ref{S:hot} (c.f.
Refs.\cite{h&w2,kap,ben}). We calculate the proportion of the
total charge that occupies the lowest energy state in
section~\ref{S:bec}, and as expected, the canonical and grand
canonical ensembles are found to give the same results in the
infinite-volume limit. In section~\ref{S:fv} we consider
finite-volume corrections where a difference between the two
ensembles is observed.


\section{Exactly conserved charge}

To study a conserved charge $Q$ using the canonical ensemble we
must identify and only average over those states with this
specific charge. As noted above, this is taken care of by
inserting a delta-function into the trace over states. The
partition function is then
    \be Z = {\rm
    Tr}\left\{\delta_{Q,\widehat{Q}} \;\widehat{\rho}\right\}.
    \ee
The density operator $\widehat{\rho}$ in the canonical ensemble
gives a probability to each state in accordance with the
temperature of the system $T=1/\beta$. It is given by
    \be \widehat{\rho}={\rm
    e}^{-\beta \widehat{H}},
    \ee
where $\widehat{H}$ is the Hamiltonian operator.

The total charge $Q$ has integer value and so the integral form of
the delta-function is
    \be
    \delta_{Q,Q'}=\frac{1}{2\pi}\int_{-\pi}^{\pi} {\rm d}\sigma\; {\rm e}^{i (Q-Q') \sigma} .
    \ee
The partition function now contains a trace over states where the
density operator is multiplied by a charge operator-dependent
phase factor
    \be
    Z = \frac{1}{2\pi}\int_{-\pi}^{\pi} {\rm
    d}\sigma\; {\rm e}^{i Q \sigma}\; {\rm Tr}\left\{ {\rm e}^{-i
    \widehat{Q} \sigma} \;\widehat{\rho}\right\}.
    \label{eq:Z4}
    \ee
This phase factor can be absorbed into the states where its effect
is felt through a change in the boundary conditions. However, it
is simpler to absorb it as a shift in the Hamiltonian of the
system:
    \be \widehat{H}\rightarrow \widehat{H}' =
    \widehat{H}+i\frac{\sigma}{\beta}\widehat{Q}. \ee
The non-hermiticity of this Hamiltonian would concern us if we
were to apply it to a dynamical problem. In thermal equilibrium
this is nothing more than a mathematical trick.

We shall consider a relativistic $O(2)$-invariant scalar theory.
This is one of the simplest theories which contains a conserved
charge. It is useful as a simplified version of the Higgs sector
and also for describing atomic gases in its non-relativistic
limit. The Hamiltonian and charge are given by
    \bea
    \widehat{H}&=&\int_V {\rm d}^3 x\; \frac{1}{2} \left[
    \widehat{\pi}_1^2+\widehat{\pi}_2^2+(\nabla \widehat{\phi}_1)^2
    +(\nabla \widehat{\phi}_2)^2+m^2 \widehat{\phi}_1^2+m^2 \widehat{\phi}_2^2 \right],\nonumber\\
    \widehat{Q}&=&\int_V {\rm d}^3 x\; (\widehat{\phi}_2\widehat{\pi}_1
    -\widehat{\phi}_1\widehat{\pi}_2).
    \eea

The trace in Eq.(\ref{eq:Z4}) can be expressed as a functional
integral in terms of the Lagrangian for this system \cite{land}
   \be
    {\rm Tr}\left\{
    {\rm e}^{-i \widehat{Q} \sigma}
    \;\widehat{\rho}\right\}=
    {\rm Tr}\left\{ {\rm e}^{-\beta\widehat{H}'} \right\} \propto
    \int [{\rm d} \phi_{a}]\;
    {\rm e}^{i\int_0^{-i\beta}{\rm d}t\;L'(t)}
    \ee
where the Lagrangian is found to be
    \bea
    L'(t)=\int_V {\rm d}^3 x \;\frac{1}{2} \left[
    \left(\partial_t \phi_1-i\frac{\sigma}{\beta}\phi_2\right)^2
    +\left(\partial_t \phi_2+i\frac{\sigma}{\beta}\phi_1\right)^2
    \right. \nonumber \\
    \left. \phantom{\left(\partial_t \phi_1-i\frac{\sigma}{\beta}\phi_2\right)^2}
    -(\nabla \phi_1)^2-(\nabla \phi_2)^2
    -m^2 \phi_1^2-m^2 \phi_2^2 \right].
    \eea
To calculate the functional integral we begin by Fourier
decomposing the fields~\cite{land}. Whilst doing this we shall
anticipate the existence of a condensate by giving the fields a
spatially constant component $v_a$ to be determined~\cite{kap}.
The constant $v_a$ acts as a classical replacement for
$\phi_{a;0}(\bf 0)$. We define $\tau=it$ where $0\le \tau \le
\beta$. The fields are periodic over this imaginary time interval
and we have
    \be
    \phi_a ({\bf x},\tau)=v_a + \frac{1}{\sqrt{\beta}}\sum_{n}\int_{\rm p}
    \;{\rm e}^{i\omega_n\tau + i{\bf p\cdot x}}\; \phi_{a;n}(\bf p),
    \ee
where $\omega_n=2\pi n/\beta$, and
    \be
    \int_{\rm p}=\int\frac{{\rm d}^3 p}{(2\pi)^3}.
    \ee
The action is now given by
    \bea i\int_0^{-i\beta}{\rm d}t\;
    L'(t)&=& -\frac{1}{2}\beta
    V\left(m^2+\frac{\sigma^2}{\beta^2}\right)\mid v\mid^2
    \nonumber\\&& +\sum_n\int_{\rm p}\frac{1}{2} \phi_{a;-n}({\bf
    -p})K_{ab;n}({\bf p})\phi_{b;n}({\bf p}), \eea
where, after defining $\omega^2={\bf p}^2+m^2$, the inverse
propagator is
    \be -K_n({\bf
    p})=\left(\begin{array}{cc}
    \omega_n^2+\omega^2+\frac{\sigma^2}{\beta^2}
    & 2i\frac{\sigma}{\beta}\omega_n \\
    -2i\frac{\sigma}{\beta}\omega_n^2 &
    \omega_n+\omega^2+\frac{\sigma^2}{\beta^2}
    \end{array} \right).
    \ee
It can be checked that this action is real. The trace in
Eq.(\ref{eq:Z4}) now becomes
    \bea
    {\rm Tr}\left\{  {\rm e}^{-i \widehat{Q} \sigma}
    \;\widehat{\rho}     \right\} &\propto&
    {\rm e}^{-\frac{1}{2}\beta V\left(m^2+\frac{\sigma^2}{\beta^2}\right)
    \mid v\mid^2}
    \int [{\rm d} \phi_{a;n}]\;{\rm e}^{\sum\int\frac{1}{2}
    \phi_{a;-n}({\bf -p})K_{ab;n}({\bf p})\phi_{b;n}({\bf p})}\nonumber\\
    &\propto&{\rm e} ^{-\frac{1}{2}\beta
    V\left(m^2+\frac{\sigma^2}{\beta^2}\right)\mid v\mid^2
    -\ln\sqrt{\det K}}. \eea
A few lines of algebra determine the integral form for
$\ln\sqrt{\det K}$. We find
    \be \ln\sqrt{\det K}=V\int_{\rm
    p}\left[\beta\omega+ \ln(1-{\rm e}^{-\beta\omega+i\sigma})+
    \ln(1-{\rm e}^{-\beta\omega-i\sigma})\right].
    \label{eq:log}
    \ee
The first term on the right-hand side of Eq.(\ref{eq:log}) is the
infinite zero-point contribution which can be factored out since
it doesn't depend on $\sigma$. The other contributions cannot be
easily calculated. We are forced to consider their behavior in
extreme-temperature limits. Our Eq.(\ref{eq:log}) is essentially
the same as Eq.(2.57) in Ref.\cite{FTFT} with the replacement
$\sigma=i\beta\mu$ where $\mu$ is the chemical potential.


\section{High-temperature and large-volume}
\label{S:hot}

Here we shall demonstrate how the contributions to $\ln\sqrt{\det
K}$ can be calculated in the high-temperature and large-volume
limit. We first perform a Taylor expansion:
    \be
    \ln(1-{\rm e}^{-\beta\omega\pm i\sigma})=-\sum_{n=1}^{\infty}
    \frac{{\rm e}^{-n\beta\omega\pm in\sigma}}{n}.
    \ee
The momentum integration can now be taken to act only on the relevant
exponential factor
    \be
    \int_{\rm p}\ln(1-{\rm e}^{-\beta\omega\pm i\sigma})=-\sum_{n=1}^{\infty}
    \frac{{\rm e}^{\pm in\sigma}}{n}\int_{\rm p}{\rm e}^{-n\beta\omega}.
    \ee
To perform this integral in the limit of small $\beta$ we first write out
the Mellin-Barnes integral form for the exponential
    \be
    {\rm e}^{-a}=\frac{1}{2\pi i}\int_{c-i\infty}^{c+i\infty}{\rm d}s\;
    a^{-s}\;\Gamma(s),
    \ee

    \be
    \int_{\rm p}{\rm e}^{-n\beta\omega}=\frac{1}{2\pi i}
    \int_{c-i\infty}^{c+i\infty}{\rm d}s\;
    (n\beta)^{-s}\;\left[\int_{\rm p}\omega^{-s}\right]\;\Gamma(s).
    \ee
The momentum integral is now straightforward
    \be
    \int_{\rm p}\omega^{-s}=\frac{m^{3-s}}{(4\pi)^{\frac{3}{2}}}
    \frac{\Gamma\left(\frac{s-3}{2}\right)}{\Gamma\left(\frac{s}{2}\right)}.
    \ee
The value of $c$ must be greater than $3$ such that the
Mellin-Barnes integrand is finite. The contour may then be closed
at infinity in the negative real half-plane where the contribution
to the integral is zero. The leading pole comes from the factor
$\Gamma\left(\frac{s-3}{2}\right)$. Evaluating its residue gives
    \be
    \int_{\rm p}{\rm e}^{-n\beta\omega}\sim
    \frac{1}{\pi^2}\left(\frac{1}{n\beta}\right)^3.
    \ee
This result could also have been found by taking $\omega \sim
p+\cdots$. This is a valid approximation when $\beta m \ll 1$.
However, the Mellin transform allows us to consider corrections to
this limit in a controlled way (see later). We therefore have
    \be
    \int_{\rm p}\ln(1-{\rm e}^{-\beta\omega\pm i\sigma})\sim-\frac{1}{\pi^2\beta^3}
    \sum_{n=1}^{\infty}
    \frac{{\rm e}^{\pm in\sigma}}{n^4}.
    \ee
Finally, summing the two logarithmic contributions to
$\ln\sqrt{\det K}$ we find \cite{GR}
    \bea
    V\int_{\rm p}\left[
    \ln(1-{\rm e}^{-\beta\omega+i\sigma})+
    \ln(1-{\rm e}^{-\beta\omega-i\sigma})\right]
    \sim -\frac{2V}{\pi^2\beta^3}
    \sum_{n=1}^{\infty}
    \frac{\cos{n\sigma}}{n^4}\nonumber\\
    =-\frac{2V}{\pi^2\beta^3}\left(\frac{\pi^4}{90}-\frac{\pi^2\sigma^2}{12}
    +\frac{\pi\mid\sigma\mid^3}{12}-\frac{\sigma^4}{48}\right).
    \label{eq:cos4} \eea

We can now bring everything together in order to write out the
canonical partition function with a conserved charge $Q$. We may
factor out any $\sigma$-independent terms as an overall
normalization $\cal N$. We also write in terms of a new variable
$x$ where $2\pi x =\sigma$. We thus have
    \be
    Z \sim {\cal N}(\beta)
    \int_{-\frac{1}{2}}^{\frac{1}{2}} {\rm d}x\; {\rm e}^{2\pi i Q x}\;
    {\rm e}^{-\frac{1}{2}\beta V\left(m^2+\frac{4\pi^2 x^2}{\beta^2}\right)
    \mid v\mid^2}\;
    {\rm e}^{-\frac{2V\pi^2}{3\beta^3}x^2(\mid x\mid -1)^2}.
    \ee
Within this range of integration, in the limit of large $V$ and
small $\beta$ we may make the approximation
    \be
    {\rm e}^{-\frac{2V\pi^2}{3\beta^3}x^2(\mid x\mid -1)^2}
    \sim {\rm e}^{-\frac{2V\pi^2}{3\beta^3}x^2}.
    \ee
Since this exponential factor suppresses any potential
contribution for larger values of $x$ we may further extend the
integration range to infinity with negligible effect giving
    \be
    Z \sim {\cal N}(\beta) \int_{-\infty}^{\infty} {\rm d}x\;
    \cos({2\pi Q x})\; {\rm e}^{-\frac{1}{2}\beta
    V\left(m^2+\frac{4\pi^2 x^2}{\beta^2}\right) \mid v\mid^2}\; {\rm
    e}^{-\frac{2V\pi^2}{3\beta^3}x^2}.
    \label{eq:zcalc}
    \ee
This is the canonical partition function for an $O(2)$-invariant
scalar field theory with total conserved charge $Q$. The integral
can be performed analytically. In the next section we shall use
$Z$ to demonstrate Bose-Einstein condensation in the canonical
ensemble.


\section{Bose-Einstein condensation}
\label{S:bec}

The classical variable $v_a$ represents the constant expectation
value of the field and is to be determined. We first identify $\ln
Z(v_a)$ with the effective potential for the field. The correct
solution $v_a$ must minimize the effective potential:
    \be
    \frac{{\rm d}}{{\rm d} v_a}\ln Z
    =\frac{1}{Z}\frac{{\rm d} Z}{{\rm d} v_a}
    =0.
    \ee
Applying this condition to $Z$ in Eq.(\ref{eq:zcalc}) we find
    \be
    0=v_a\left(m^2+\frac{4\pi^2}{\beta^2}\frac{\int_0^{\infty}
    {\rm d}x\; \cos({2\pi Q x})\;x^2\;
    {\rm e}^{-\frac{2V\pi^2}{\beta}
    \left(\frac{1}{3\beta^2}+{\mid v\mid^2}\right)x^2}}
    {\int_0^{\infty}
    {\rm d}x\; \cos({2\pi Q x})\;
    {\rm e}^{-\frac{2V\pi^2}{\beta}
    \left(\frac{1}{3\beta^2}+{\mid v\mid^2}\right)
    x^2}}
    \right).
    \ee
This can be evaluated with the following standard integrals
\cite{GR}
    \bea
    \int_0^{\infty}{\rm d}x\;\cos(ax)\;{\rm e}^{-bx^2}
    &=&\frac{1}{2}\sqrt{\frac{\pi}{b}}\;{\rm e}^{-\frac{a^2}{4b}},
    \\
    \int_0^{\infty}{\rm d}x\;\cos(ax)\;x^2\;{\rm e}^{-bx^2}
    &=&\frac{1}{2}\sqrt{\frac{\pi}{b}}\;
    \left(\frac{1}{2b}-\frac{a^2}{4b^2}\right)\;
    {\rm e}^{-\frac{a^2}{4b}}.
    \eea
Given these integrals we find
    \be
    0=v_a\left(m^2-\frac{Q^2}{V^2\left(\frac{1}{3\beta^2}+\mid v\mid^2
    \right)^2}\right).
    \ee
This constraint equation looks very much like the grand canonical
constraint (see Eq.(1.42) in \cite{kap}) once we define the
chemical potential by
    \be
    \mu=\frac{Q}{V\left(\frac{1}{3\beta^2}+\mid v\mid^2\right)}.
    \ee
There are two solutions:
    \bea
    v_a &=& 0\\
    \mid v\mid^2 &=& \frac{Q}{Vm}-\frac{1}{3\beta^2}.
    \eea
The first solution is trivial. The second solution exists when
$\frac{Q}{Vm}>\frac{1}{3\beta^2}$. We may therefore define a
critical temperature to mark the point at which the field develops
a non-zero expectation value
    \be
    T_c=\frac{1}{\beta_c}=\sqrt{\frac{3Q}{Vm}}.
    \ee

To express the charge stored in the ground state we write out the
total charge in integral form:
    \be
    Q=\langle Q\rangle =
    \frac{1}{2\pi Z}
    \int_{-\pi}^{\pi} {\rm d}\sigma\; {\rm e}^{i Q \sigma}\;
    i\frac{\rm d}{{\rm d}\sigma}
    {\rm Tr}\left\{{\rm e}^{-i \widehat{Q} \sigma}
    \;\widehat{\rho}\right\}
    \ee
giving
    \bea
    Q&=&\frac{2V\pi}{\beta}\left(\frac{1}{3\beta^2}+{\mid v\mid^2}\right)
    \frac{\int_0^{\infty}
    {\rm d}x\; \sin({2\pi Q x})\;x\;
    {\rm e}^{-\frac{2V\pi^2}{\beta}
    \left(\frac{1}{3\beta^2}+{\mid v\mid^2}\right)x^2}}
    {\int_0^{\infty}
    {\rm d}x\; \cos({2\pi Q x})\;
    {\rm e}^{-\frac{2V\pi^2}{\beta}
    \left(\frac{1}{3\beta^2}+{\mid v\mid^2}\right)
    x^2}}\\
    &=&Q_{\rm fluct}+Q_{\rm cond}.
    \eea
Here we can divide the contributions to the charge into two types.
The first term in brackets corresponds to the charge contained in
the fluctuations or excited modes. The second term is the charge
which is contained in the ground state, the Bose-Einstein
condensate. We find
    \be
    Q_{\rm cond}=\mid v\mid^2
    \frac{Q}{\left(\frac{1}{3\beta^2}+\mid v\mid^2\right)}
    =\left\{\begin{array}{ll}
    0 & \textrm{when $\beta^2<\frac{Vm}{3Q}$}\\
    Q-\frac{Vm}{3\beta^2} & \textrm{when $\beta^2>\frac{Vm}{3Q}$}
    \end{array} \right. .
    \ee
At $T=1/\beta=0$ all the charge resides in the condensate. As the
temperature increases charge is excited out. Eventually, when
$T>T_c$, the condensate melts and all the charge resides in the
fluctuation modes. Of course, these equations are only valid in
the high-temperature limit and we cannot rely on them when the
condensate becomes big.

It has been demonstrated how we can understand BEC without using a
chemical potential. We achieve the same result as for the grand
canonical ensemble in the high-temperature, large-volume limit.


\section{Finite-volume}
\label{S:fv}

Up until now, the canonical ensemble has proved to be more
complicated than the grand canonical ensemble without offering any
new results. However, now we stand to benefit from the fact that
the canonical ensemble provides a more realistic picture of a
finite-volume system, closed to the movement of charge into and
out-of that volume. To continue from the previous section we work
in the high-temperature region and the bosons are not subjected to
any external potential within their volume.

It is well known that there is no phase transition involved in
Bose-Einstein condensation for finite volumes. However, we may
still characterize the process by studying the proportion of the
total charge stored in the lowest energy state. In the grand
canonical ensemble the charge in the lowest energy state is given
by
    \be
    Q_0=\frac{2T\mu}{m^2-\mu^2}.
    \ee
For $T<T_c$ when the charge occupying the lowest energy state
compares with the charge stored in fluctuations, we may think of
$(m-\mu)^{-1} \sim {\cal O}(V m T)$ \cite{toms}. The chemical
potential is determined by the constraint that the total energy is
fixed. To leading order in the large-volume expansion we have
\cite{H&W}
    \be
    Q=\frac{2T\mu}{m^2-\mu^2}+V\left[\frac{\mu T^2}{3}
    +\frac{\mu T (m^2-\mu^2)^{\frac{1}{2}}}{2\pi}
    +\frac{\mu(3m^2-2\mu^2)}{12\pi^2}+\cdots\right]
    +\cdots
    \label{eq:gc}
    \ee
For any given $T$, $V$ and $Q$, this equation can be solved for
$\mu$ enabling $Q_0$ to be calculated. (For a more detailed
finite-volume calculation see \cite{HU}.)

Returning to the canonical ensemble and the expression for
$\ln\sqrt{\det K}$ given in equation (\ref{eq:log}), for finite
volumes, the integral over momentum states should be replaced with
an infinite sum over those momentum states compatible with the
boundary conditions. We again expand as follows
    \be
    \sum_{\rm p}\ln(1-{\rm  e}^{-\beta\omega\pm i\sigma})=-\sum_{n=1}^{\infty}
    \frac{{\rm e}^{\pm in\sigma}}{n}\sum_{\rm p}{\rm
    e}^{-n\beta\omega},
    \ee

    \be
    \sum_{\rm p}{\rm e}^{-n\beta\omega}=\frac{1}{2\pi i}
    \int_{c-i\infty}^{c+i\infty}{\rm d}s\;
    (n\beta)^{-s}\;\left[\sum_{\rm p}\omega^{-s}\right]\;\Gamma(s).
    \ee
Choosing a cubic container for our system with side $L$ and
Neumann boundary conditions we have
    \bea
    \sum_{\rm p}\omega^{-s}&=&\sum_{\rm p}\left[
        {\rm p}^2+m^2 \right]^{-\frac{s}{2}}\nonumber\\
    &=&\sum_{n_i=0}^{\infty}\left[
        \left(\frac{\pi n_1}{L}\right)^2
        + \left(\frac{\pi n_1}{L}\right)^2
        + \left(\frac{\pi n_1}{L}\right)^2
               +m^2 \right]^{-\frac{s}{2}}\nonumber\\
    &=&\frac{1}{\Gamma\left(\frac{s}{2}\right)}\int_0^{\infty}
    {\rm d} t \; t^{\frac{s}{2}-1} {\rm e}^{-m^2 t}
    \sum_{\rm p}{\rm e}^{-{\rm p}^2 t}.
    \eea
The sum over $\rm p$ can be performed as a large-volume expansion
\cite{HU}
    \bea
    \sum_{\rm p}{\rm e}^{-{\rm p}^2 t}
    &=&\left[\sum_{n=0}^{\infty}{\rm e}^{-(\frac{\pi n}{L})^2
    t}\right]^3\nonumber\\
    &=& \frac{V}{8 (\pi t)^{\frac{3}{2}}}+\frac{3V^{\frac{2}{3}}}{8 \pi t}
        +\frac{3V^{\frac{1}{3}}}{8 (\pi t)^{\frac{1}{2}}}+\frac{1}{8}+\cdots
    \eea
This gives
    \bea
    \sum_{\rm p}\omega^{-s}&=&V\frac{
    m^{3-s}\Gamma\left(\frac{s-3}{2}\right)}
    {(4\pi)^{\frac{3}{2}}\Gamma\left(\frac{s}{2}\right)}
    +V^{\frac{2}{3}}\frac{
    3m^{2-s}\Gamma\left(\frac{s-2}{2}\right)}
    {8\pi\Gamma\left(\frac{s}{2}\right)}
    \nonumber\\&&+V^{\frac{1}{3}}\frac{
    3m^{1-s}\Gamma\left(\frac{s-1}{2}\right)}
    {8\pi^{\frac{1}{2}}\Gamma\left(\frac{s}{2}\right)}
    +\frac{m^{-s}}{8}+\cdots,
    \eea
and eventually
    \bea
    \ln\sqrt{\det K} &\sim&
    -V\frac{2}{\pi^2\beta^3}\sum_{n=1}^{\infty}\frac{\cos(n\sigma)}{n^4}
    +Vm^2\frac{1}{2\pi^2\beta}\sum_{n=1}^{\infty}\frac{\cos(n\sigma)}{n^2}
    \nonumber\\
    &&-V^{\frac{2}{3}}\frac{3}{2\pi\beta^2}\sum_{n=1}^{\infty}\frac{\cos(n\sigma)}{n^3}
    +V^{\frac{2}{3}}m^2\frac{3}{4\pi}\sum_{n=1}^{\infty}\frac{\cos(n\sigma)}{n}
    \nonumber\\
    &&-V^{\frac{1}{3}}\frac{3}{2\pi\beta}\sum_{n=1}^{\infty}\frac{\cos(n\sigma)}{n^2}
    -2\sum_{n=1}^{\infty}\frac{{\rm e}^{-n\beta m}\cos(n\sigma)}{n}
    +\cdots.
    \label{eq:LDK}
    \eea
We have kept all terms in the $1/L$ expansion up to ${\cal O}(L^0)
$ and in the high temperature (small mass) expansion up to ${\cal
O}(m^2)$. All but one of these infinite sums have analytic
solutions for the given integration range of $\sigma$. Referring
to~\cite{GR} we have
    \bea
    \sum_{n=1}^{\infty}\frac{\cos(n\sigma)}{n^4}
    &=&\frac{\pi^4}{90}-\frac{\pi^2\sigma^2}{12}
    +\frac{\pi|\sigma|^3}{12}-\frac{\sigma^4}{48}\\
    \sum_{n=1}^{\infty}\frac{\cos(n\sigma)}{n^2}
    &=&\frac{\pi^2}{6}-\frac{\pi|\sigma|}{2}+\frac{\sigma^2}{4},\\
    \sum_{n=1}^{\infty}\frac{\cos(n\sigma)}{n}
    &=&-\frac{1}{2}\ln(2-2\cos\sigma),\\
    \sum_{n=1}^{\infty}\frac{{\rm e}^{-n\beta m}\cos(n\sigma)}{n}
    &=&-\frac{1}{2}\left[\ln(1-{\rm e}^{-\beta m+i\sigma})
        +\ln(1-{\rm e}^{-\beta m-i\sigma})\right].
    \eea
The remaining sum is convergent and can easily be evaluated
numerically.

There is no guarantee that the expansion of Eq.(\ref{eq:LDK}) is a
good one. The best we can expect is that subsequent terms in the
expansion quickly become negligibly small in comparison with the
leading order terms over the whole integration range of $\sigma$.
This is clearly not the case for the fourth term in the expansion
when close to $\sigma=0$. The term diverges leading to a zero in
the integrand. However, by appropriate choice of parameters, the
domination of this term is made sufficiently brief that its effect
is negligible.

The ${\cal O}(V^0)$ term of Eq.(\ref{eq:LDK}) may be rewritten as
    \be
    \frac{1}{2}\sum_{n}\ln\left[(\omega_n^2+m^2+T^2\sigma^2)^2-T^2\sigma^2\omega_n^2\right]
    \ee
from which we may extract the $n=0$ contribution and identify this
as the contribution to $\ln\sqrt{\det K}$ of the ground state mode
$\phi_{a;0}(\bf 0)$:
    \be
    \ln(m^2+T^2\sigma^2).
    \label{eq:gsc}
    \ee
The expectation of charge in the lowest energy state is given by
    \be
    \langle Q_0 \rangle = \frac{1}{2\pi Z}\int_{-\pi}^{\pi} {\rm
    d}\sigma\; {\rm e}^{i Q \sigma}\; {\rm Tr}\left\{{\rm e}^{-i
    \widehat{Q} \sigma} \;\widehat{\rho}\;\widehat{Q}_0 \right\}.
    \ee
This can be derived by differentiating with respect to $\sigma$
only those terms in $Z$ where $\sigma$ is coupled to the ground
state mode (expression~(\ref{eq:gsc})). We find
    \be
    \langle Q_0 \rangle = \frac{1}{2\pi Z}\int_{-\pi}^{\pi} {\rm
    d}\sigma\; {\rm e}^{i Q \sigma}\;\left(\frac{-2i\sigma}
    {\beta^2 m^2+\sigma^2}\right)
    {\rm e} ^{-\ln\sqrt{\det K}}
    \ee
where $\ln\sqrt{\det K}$ is given above in (\ref{eq:LDK}).

\begin{figure}[t]
    \epsfxsize=18pc
    \[\epsfbox{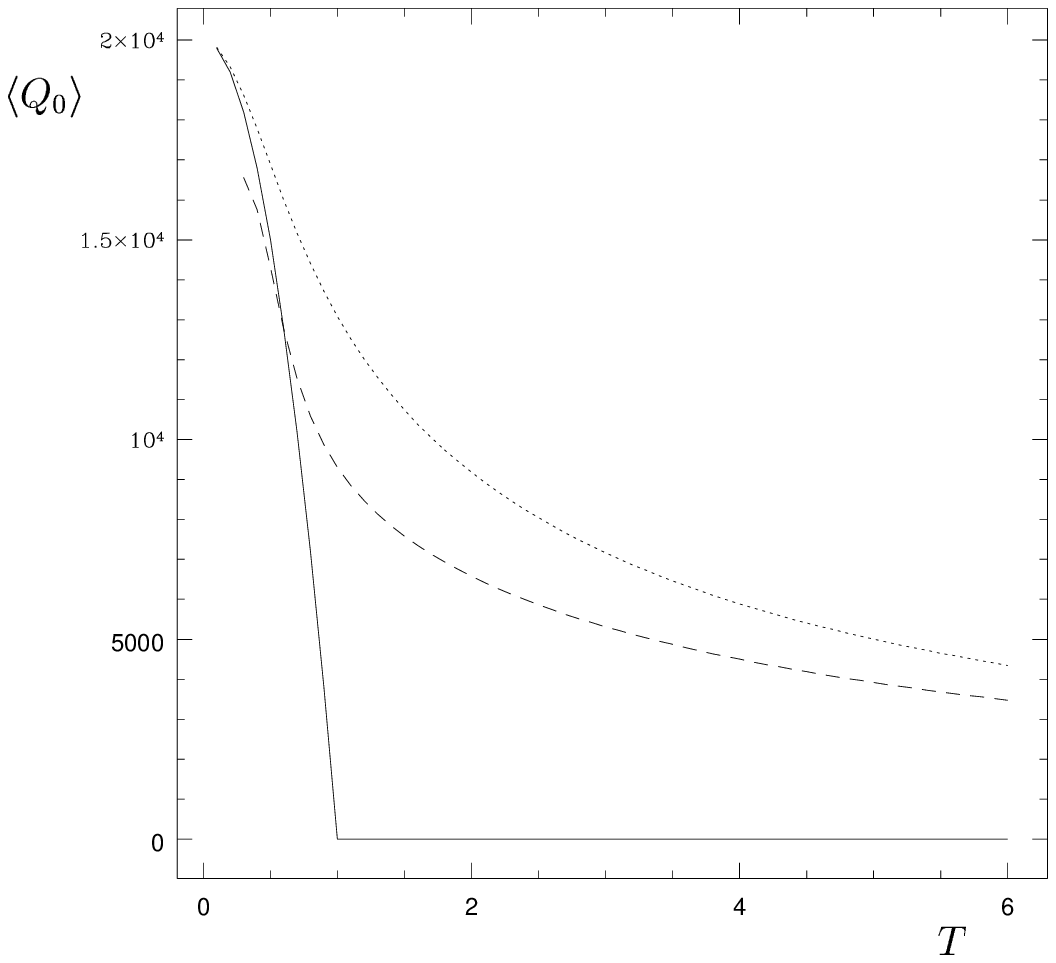}\]
    \caption{\small Expectations of charge in the lowest energy state
    against temperature. The solid line represents the infinite-volume
    approximation in the high-temperature limit; the dotted line is a
    finite-volume approximation using the grand canonical ensemble; the
    dashed line uses the canonical ensemble. Parameters are $Q=20000$,
    $V=10^9$, mass is chosen to give $T_c=1$ ($m=0.00006$). } \label{fig:bec1}
    \end{figure}

\begin{figure}[t]
    \epsfxsize=18pc
    \[\epsfbox{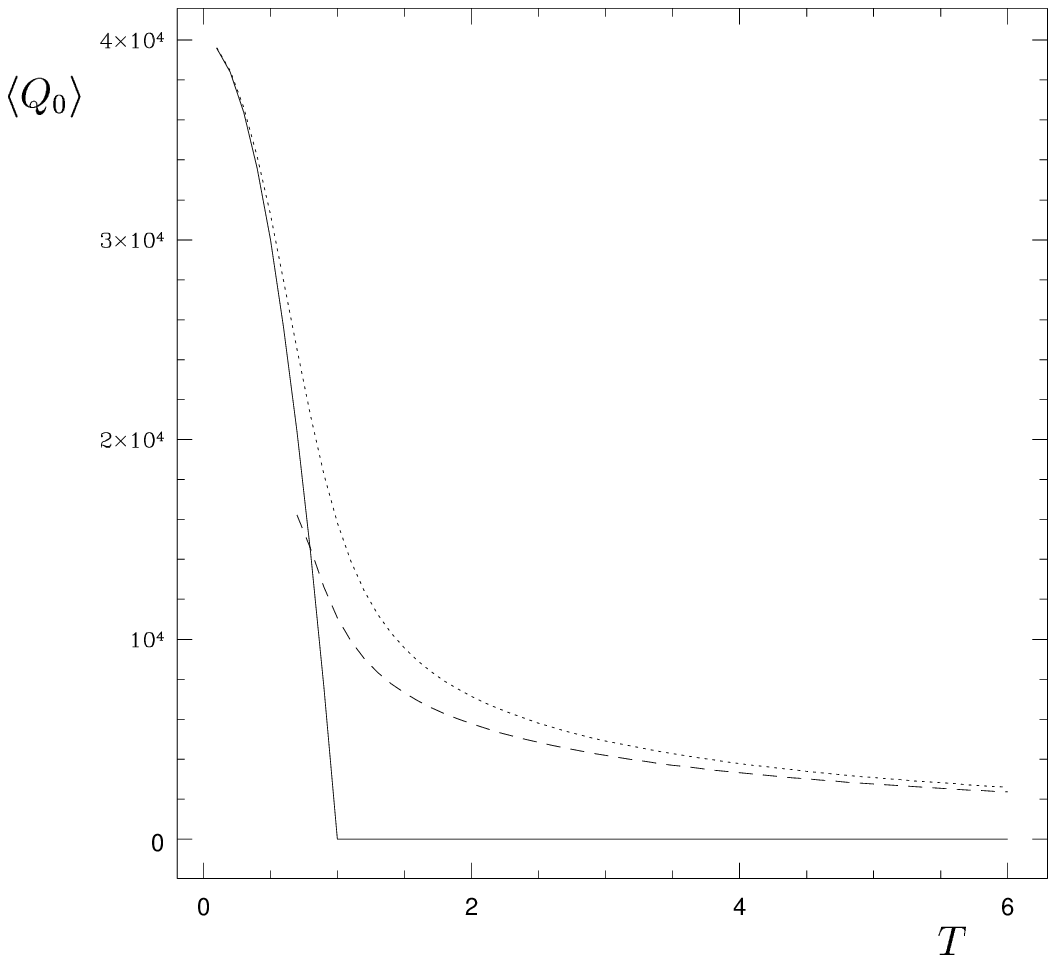}\]
    \caption{\small $Q=40000$,
    $V=10^9$, $m=0.00012$ in units of $T_c$.
    Canonical (dashed line), grand canonical (dotted line),
    infinite volume approximation (solid line).
    } \label{fig:bec2}
    \end{figure}

Having defined the condensate by the expectation of charge in the
lowest energy mode, we may directly calculate the condensate
fraction by numerical integration. The results can be seen for
different values of the total charge in Fig.\ref{fig:bec1} and
Fig.\ref{fig:bec2}. A comparison is made with the infinite-volume
limit and with the finite-volume approximation in the grand
canonical ensemble given by Eq.(\ref{eq:gc}). The parameters must
be chosen such that $\beta/V^{1/3}, \beta m \ll 1$ in order to
justify the expansion of $\ln\sqrt{\det K}$. We chose $Q\sim10^4$,
$V=10^9/T_c^3$ and $m=3Q/(VT_c^2)$, working in units where
$T_c=1$.

For $T>T_c$ we consistently observe a lower occupation of the
ground state in the canonical ensemble as compared with the grand
canonical ensemble. This makes the infinite-volume limit a better
approximation for the canonical than the grand canonical case. As
the temperature approaches the critical temperature we see the
ground state occupation increase as expected. At low temperatures
the expansion of $\ln\sqrt{\det K}$ breaks down and canonical
results deviate from the other approximations.

\section{Conclusions}

It has been shown how we can understand BEC for a system of fixed
charge in the canonical ensemble. Though the chemical potential
plays a crucial role in our understanding of BEC in the grand
canonical ensemble, there is no need for a chemical potential in
the canonical ensemble.

We have seen that the two ensembles produce identical results in
the infinite-volume limit. For finite volumes, the canonical
method produces sensible results which differ from the grand
canonical approximation in a consistent qualitative way.

For future work, a low-temperature, non-relativistic version of
this calculation could feasibly be tested against experimental
data from BEC in atomic fluids and other theoretical results
\cite{poli,herz}.


\section*{Acknowledgements}

This work was financially supported by The Royal Commission for
the Exhibition of 1851. I would like to acknowledge Tim Evans and
Marko Ivin for discussions concerning the canonical ensemble. I
would also like to acknowledge the hospitality of the particle
theory group at the University of Sussex and the HEP group at UCL.


\end{document}